\UseRawInputEncoding
\documentclass[journal=cmatex,manuscript=article]{achemso}
\mciteErrorOnUnknownfalse
\usepackage{graphicx}
\usepackage{amsmath}
\usepackage{float}
\usepackage{svg}
\usepackage{dirtytalk}
\usepackage{multirow}
\usepackage{booktabs,makecell}
\usepackage{siunitx}
\usepackage{chemformula}
\usepackage{tablefootnote}
\usepackage{mleftright}
\usepackage{soul,color}
\usepackage{threeparttable}

\graphicspath{ {Images/} }

\newcounter{daggerfootnote}

\SectionNumbersOn
\author{Rishi Gurnani}
\author{Christopher Kuenneth}
\author{Aubrey Toland}
\author{Rampi Ramprasad}\email{rampi.ramprasad@mse.gatech.edu}
\affiliation[MSE]{School of Materials Science and Engineering, Georgia Institute of Technology, 30332 Atlanta, Georgia, United States}

\title{Polymer informatics at-scale with multitask graph neural networks}

\begin{document}

\begin{abstract}
Artificial intelligence-based methods are becoming increasingly effective at screening libraries of polymers down to a selection that is manageable for experimental inquiry. The vast majority of presently adopted approaches for polymer screening rely on handcrafted chemostructural features extracted from polymer repeat units\textemdash a burdensome task as polymer libraries, which approximate the polymer chemical search space, progressively grow over time. Here, we demonstrate that directly ``machine-learning'' important features from a polymer repeat unit is a cheap and viable alternative to extracting expensive features by hand. Our approach\textemdash based on graph neural networks, multitask learning, and other advanced deep learning techniques\textemdash speeds up feature extraction by one to two orders of magnitude relative to presently adopted handcrafted methods without compromising model accuracy for a variety of polymer property prediction tasks. We anticipate that our approach, which unlocks the screening of truly massive polymer libraries at scale, will enable more sophisticated and large scale screening technologies in the field of polymer informatics.
\end{abstract}

\section{Introduction}

Polymers have emerged as a powerful class of materials for a wide range of applications because of their low-cost processing, chemical stability, tunable chemistries, and low densities. These attributes have led to vigorous, widespread, and sustained research, and to the development of new polymeric materials \cite{Baldwin2015PolydimethyltinApplications, MANNODIKANAKKITHODI2018785, Hu2022Machine-Learning-AssistedPolymers}. The result is a constant flux of materials data. Over the past decade, the polymer informatics community has translated this data stream into machine-learned property predictors that efficiently screen libraries of candidate polymers for subsequent experimental inquiry \cite{DoanTran2020Machine-learningGenome, Kuenneth2021PolymerLearning}.

Currently, most approaches for polymer screening rely on handcrafted features\textemdash extracted from the chemical structure of a polymer repeat unit\textemdash as input for property predictors \cite{Barnett2020, Patel2022FeaturizationLearning}. These approaches are highly accurate, but feature extraction becomes a bottleneck (as discussed in Section \ref{sec:speed}) when used to screen vast swathes of the polymer chemical space. This bottleneck is increasingly exposed by the proliferation of enumeration methods \cite{Ma2020PI1M:Informatics, Ruddigkeit2012EnumerationGDB-17} and long-sought \cite{Franceschetti1999TheProperties, Batra2020} inverse predictors \cite{Gurnani_2021, Batra2020PolymersAutoencoders, Kim2021PolymerLearning,Yao2021InverseModels, Zunger2018InverseFunctionalities}, which directly locate optimal pockets of the chemical space from a user-defined wish list of material properties. By leveraging these tools, the day that we routinely generate billions of polymer candidates is fast approaching. Advances in polymer screening and feature engineering are needed to keep up with this pace.

An alternative to handcrafting features is ``machine learn'' them. One approach is to represent the material as raw text, such as a simplified molecular-input line-entry system (SMILES) \cite{Weininger2002SMILESRules} or BigSMILES \cite{Lin2019BigSMILES:Macromolecules} string, and then learn features with a neural network specifically designed for natural language processing \cite{Chen2021PredictingModel}. Another promising approach is to represent the material as a graph, and then train a Graph Neural Network (GNN) \cite{Gilmer2017NeuralChemistryb} to learn features. To date, GNNs have outperformed approaches based on handcrafted features \cite{Gilmer2017NeuralChemistryb, Schutt2017SchNet:Interactions, Jrgensen2018NeuralMaterials, Hy2018PredictingNetworks, Zhang2020MolecularStructures} on the massive QM9 database \cite{Ramakrishnan2014QuantumMolecules} for small molecules. Similarly, feature learning approaches have supplanted traditional methods in other domains (e.g., convolutional neural networks \cite{Lecun2015DeepLearning} in computer vision and transformers \cite{Vaswani2017AttentionNeed} in natural language processing) where the extraction of handcrafted representations from the input data is non-trivial or impractical \cite{Lecun2015DeepLearning}.

Another important emerging trend in machine learning (ML) for materials science is multitask learning \cite{Caruana1997MultitaskLearning, Kuenneth2021PolymerLearning}. The core idea behind multitask learning is that, by training a model to simultaneously learn multiple correlated target properties, the model is less likely to produce overfitted predictions to the training set of any one target property \cite{Caruana1997MultitaskLearning}. As a result, the predictive performance for each property is improved. This idea is seen in nature as well. For example, there is evidence that training in one sport can improve a young athlete in another related sport \cite{Brenner2016SportsAthletes}.

A handful of polymer GNNs have been explored in the past \cite{Jrgensen2018MachineCells, Zeng2018GraphPredictionb, StJohn2019Message-passingScreening, Hatakeyama-Sato2020AI-AssistedStructures, Mohapatra2022Chemistry-informedLearning, Aldeghi2022APrediction, Antoniuk2022RepresentingPredictions}. The majority of these approaches are single task. The GNN proposed by Mohapatra et al.  \cite{Mohapatra2022Chemistry-informedLearning} is suitable for biopolymers, in which the monomer sequence is known. Other approaches \cite{StJohn2019Message-passingScreening, Hatakeyama-Sato2020AI-AssistedStructures, Aldeghi2022APrediction, Antoniuk2022RepresentingPredictions}, geared toward synthetic polymers (the subject of interest in this work), represent a polymer using the graph of a predominant repeat unit. This introduces the need for invariance to certain transformations of the repeat unit graph: translation, addition, and subtraction (as defined in Section \ref{sec:polyGNN}). A subset of the GNNs  for synthetic polymers \cite{Aldeghi2022APrediction, Antoniuk2022RepresentingPredictions} are invariant to translation, but not to addition and subtraction. In other words, a GNN that preserves the invariant properties of polymer repeat units has not been developed until now. Our work, a powerful multitask GNN architecture (see Fig. 1) for polymers, fills this gap. We call this development the Polymer Graph Neural Network (polyGNN).

\begin{figure*}
    \centering
	\includegraphics[width=0.99\textwidth]{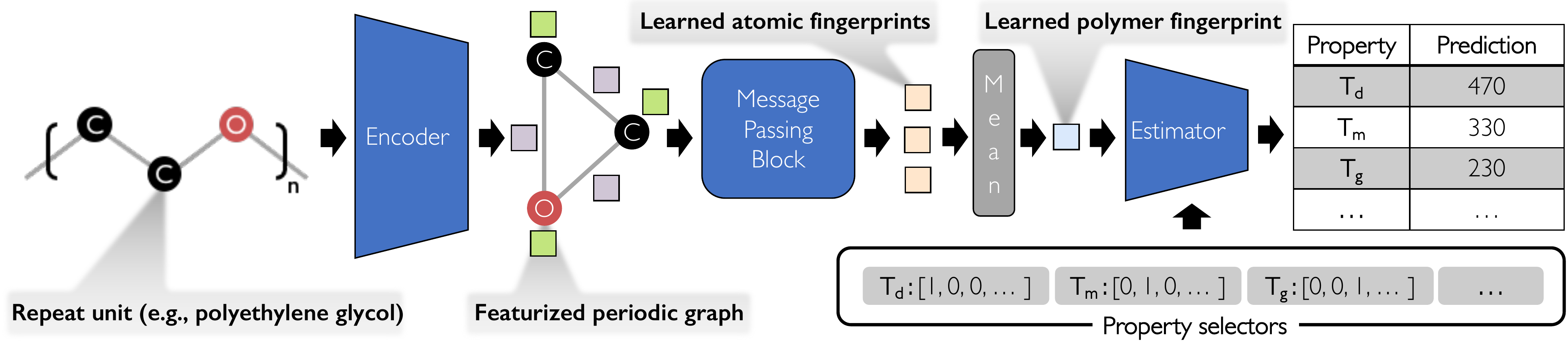}
	\caption{The polyGNN architecture. The Encoder converts the repeat unit SMILES string to a periodic graph and then computes \emph{initial} atomic and bond fingerprint vectors (green and purple squares, respectively). A subsequent set of atomic fingerprints (yellow squares) are learned by the Message Passing Block and then averaged, yielding the learned polymer fingerprint (light blue square). This fingerprint and a series of selector vectors are passed to the Estimator, producing a series of property predictions. $T_\text{d}, T_\text{m}, T_\text{g}$ refer to the critical temperatures for thermal decomposition, melting, and glass transition, respectively.}
	\label{fig:Architecture}
\end{figure*}

In the small molecule domain, the adoption of GNNs is motivated by systematic work \cite{Gilmer2017NeuralChemistryb} comparing GNNs and handcrafted approaches on even footing across a diverse set of molecules and predictive tasks. Analogous studies are absent from the synthetic polymer domain. Previous works have compared feature learning and handcrafted approaches for up to two \cite{Zeng2018GraphPredictionb, Aldeghi2022APrediction} polymer properties, or for several properties in the same class \cite{Jrgensen2018MachineCells} (e.g., electronic properties). In what follows, we compare polyGNN with the handcrafted fingerprint originally hosted under the Polymer Genome (PG) project \cite{DoanTran2020Machine-learningGenome} on a large and diverse data set consisting of more than 13,000 polymers and 30+ predictive tasks\textemdash spanning thermal, thermodynamic, physical, electronic, optical, dielectric, and mechanical properties, the Hildebrand solubility parameter, as well as permeability of six gases.

Our benchmark, the PG fingerprint, contains descriptors that correspond to one of three length scales. The finest-level components are atomic triples (e.g., C$_i$O$_j$N$_k$) where the subscripts denote the atomic coordination. The next (block) level contains pre-defined atomic fragments (e.g., the common cycloalkenes). These two levels contain strictly one-hot features. At the highest (chain) level are numerical features that describe the atomic or block topology (e.g., the number of atoms in the largest side chain). The handcrafted PG fingerprint is the current state-of-the-art in polymer representation, and has shown success in the numerical representation of materials over a wide chemical and property space \cite{Gurnani2021InterpretableFrameworks, DoanTran2020Machine-learningGenome, Ma2020PI1M:Informatics}. The handcrafted PG fingerprint-based property predictors thus serve as veritable performance baselines. We find that polyGNN, relative to these baselines, leads to a one to two orders of magnitude faster fingerprinting and better or comparable model accuracy in most polymer property prediction tasks. polyGNN thus offers a powerful new polymer informatics option for screening the polymer chemical space at scale. 

\section{Methods}
\subsection{Data set and preparation}\label{section:data}
Our corpus contains measurements for up to 36 properties of 13,388 polymers, yielding over 21,000 data points in total. The unit and symbol for each property is listed in Fig. 2A. The distribution of data points per property is plotted in Fig. 2B. These data points come from in-house density functional theory (DFT) computations \cite{Huan2015AcceleratedFingerprints, Huan2016ADesign, Sharma2014RationalDielectrics}, experimental data collected from the literature \cite{Park1997CorrelationMethod, Kim2018PolymerPredictions, Zhu2020PolymerPolymers, Chen2020Frequency-dependentLearning, Lightstone2020RefractiveLearning, Venkatram2020PredictingLearning}, printed handbooks \cite{PH1, Barton2013CRCParameters, Bicerano2002PredictionProperties}, and online databases \cite{polyinfo, Crow1}. Band gaps were calculated for both individual polymer chains $E_{\text{gc}}$ and polymer crystal (bulk) structures $E_{\text{gb}}$ using DFT. DFT data contain uncertainties due to the choice of exchange correlation functional, pseudopotentials, optimization procedure, etc. while data from physical experiment comes with uncertainty due to sample and measurement conditions. Thus, data for the same property but from different sources (e.g., DFT-computed and experimentally measured refractive index) are treated separately and then co-learned with multitask learning.

\begin{figure*}[h!]
    \centering
	\includegraphics[width=0.99\textwidth]{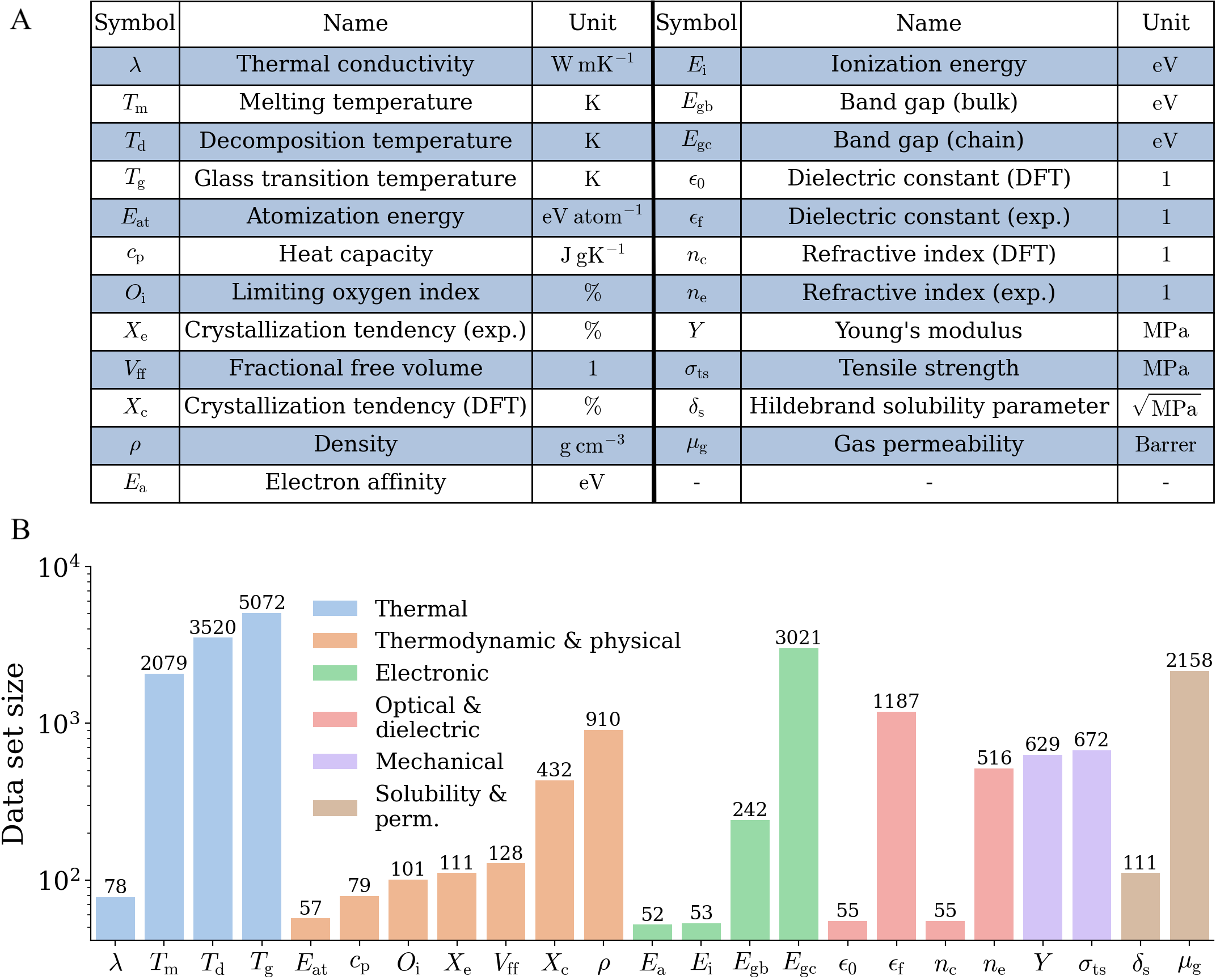}
	\caption{
    	Breakdown of our data set. \textbf{(A)} The symbol, name, and unit of each property in our data set. For properties with data from both experiment and DFT calculations, the two sources are distinguished by the abbreviations ``expt.'' and ``DFT''. Our data set includes the permeability $\mu_\text{g}$ of six gases $g \in$ \{\ch{He}, \ch{H2}, \ch{CO2}, \ch{O_2}, \ch{N_2}, \ch{CH_4}\}. Each permeability data point is scaled by $x \to \log_{10}(x+1)$. Our experimental dielectric constant $\epsilon_{\text{f}}$ data contains measurements at nine frequencies $f \in \{1.78, 2, 3, 4, 5, 6, 7, 9, 15\}$ in $\text{log}_{10}$Hz. The distribution of $\mu_\text{g}$ and $\epsilon_{\text{f}}$ are given in Section S1. \textbf{(B)} The data set size per property, shown on both the $y$-axis and above each bar. Bars of the same color belong to the same property class. ``perm.'' stands for gas permeability.
	}
	\label{fig:dataset}
\end{figure*}

Our multitask learning approach requires data preprocessing steps. First, the training data for each property was MinMax scaled between zero and one. This ensures that the optimizer of a multitask ML model equally weights the loss for each property during training. Second, to better exploit correlations between properties \cite{Kuenneth2021PolymerLearning}, we divided our entire 36 property data set into six ``property groups'': thermal properties, thermodynamic \& physical properties, electronic properties, optical \& dielectric properties, solubility \& gas permeability, and mechanical properties.The stratification of properties by group is shown in Fig. 2B. Finally, we designate each property within one group a unique one hot ``selector'' vector (see Fig. 1 for example selector vectors of thermal properties). These vectors are used by our ML models to distinguish between multiple tasks.

\subsection{polyGNN}\label{sec:polyGNN}
All GNNs rely on a well-defined graph representation of their input. If the input is a small molecule, then building a corresponding graph is straight-forward\textemdash each heavy (i.e., non-hydrogen) atom is a graph node and each bond between heavy atoms is a graph edge. However, polymers are macromolecules with numerous atoms and bonds. Creating a node and edge for each atom or bond will generate a massive graph. Machine learning based on thousands of such graphs would be computationally inefficient. Instead, we construct a polymer graph from its repeat unit alone and propose that additional information (e.g., molecular weight, end groups, etc.)\textemdash if available\textemdash be concatenated to each computed atom or bond fingerprint and/or to the learned polymer fingerprint.

Ideally, our learned polymer fingerprint must respect the invariances present in a polymer repeat unit. We identify three key transformations\textemdash translation, addition, and subtraction\textemdash that repeat units of infinite 2D polymer chains are invariant to. We define translation as the movement of the periodicity window, which can produce periodic repeat units that are all equivalent. For example, \ch{(-OCC-)}, \ch{(-COC-)}, and \ch{(-CCO-)} are equivalent repeat units of polyethylene glycol, related to one another by translation. We define addition (subtraction) as the extension (reduction) of a repeat unit by one or more minimal repeat units. For example, \ch{(-COCO-)} and \ch{(-COCOCO-)} are equivalent repeat units, related to one another by the addition (or subtraction) of their minimal repeat unit, \ch{(-CO-)}. We have constructed polyGNN to be invariant under such transformations, as discussed below.

The polyGNN architecture is composed of three main modules: an Encoder for processing the repeat unit, a Message Passing Block for fingerprinting, and an Estimator to co-learn multiple properties. In the polyGNN Encoder, bonds are added between heavy atoms at the boundary of any input repeat unit, forming a periodic polymer graph (as shown in Fig. 1). This ensures that the graph of the repeat unit, and hence its learned fingerprint, is invariant to translation. Then, each atom and bond in the graph are given initial feature vectors (described later in Section \ref{sec:graph_fp}) that are computed using RDKit \cite{RDKit1}. The featurized graph is passed to the Message Passing Block and then to the aggregation function. In the Message Passing Block, the initial feature vectors are passed between neighboring atoms. This information flow is the mechanism by which rich polymer features are learned (described later in Section \ref{sec:passing}).

After message passing, the sequence of learned atomic fingerprints is aggregated into a single polymer fingerprint by taking the \texttt{mean}. Taking the \texttt{mean} rather than the \texttt{sum} ensures that, for example, \ch{(-COCO-)} and \ch{(-COCOCO-)} are mapped to the same fingerprint. However, there are polymers (see Fig. 3A) where the desired invariance is not preserved. These conflicts arise because RDKit treats periodic polymer graphs as cyclic molecules. To address these conflicts, we propose two approaches. In the first approach, which we will continue to refer to as polyGNN, the original training data set is augmented with transformed repeat units (see Fig. 3B). Thus, although polyGNN is not invariant to addition or subtraction in these complicated cases, it achieves approximate invariance after learning from augmented data. This choice was inspired by state-of-the art image classification models, which are trained using cropped and flipped images \cite{He2016DeepRecognition}. In this work, we find that data augmentation is also effective for training polyGNNs but does increase training time\textemdash a one-time cost. As an alternative, we created a variant, polyGNN2, with guaranteed invariance to addition and subtraction (and thus no need for augmentation). Invariance is achieved by modifying the Encoder to compute features on an extended polymer graph instead of on the periodic graph (see Section S2). However, operating on the extended graph notably slows fingerprinting in polyGNN2, and so we instead focus on polyGNN in what follows.

\begin{figure}[ht]
    \centering
	\includegraphics[width=0.99\textwidth]{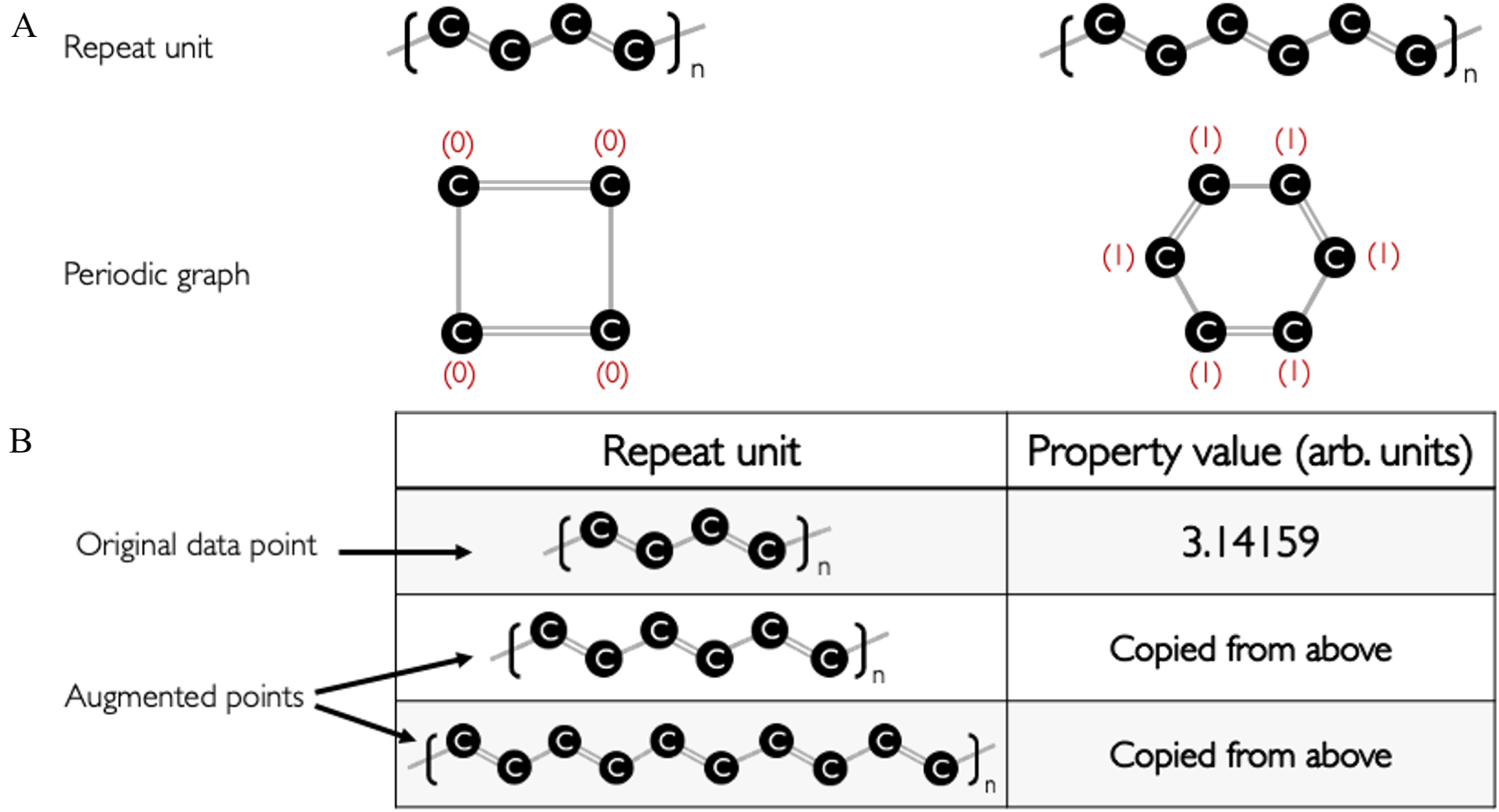}
	\caption{Overview of data set augmentation. \textbf{(A)} Two equivalent repeat units of infinite polyacetylene and their corresponding periodic graphs. Each atom in the graph is labeled with a zero if the atom is aliphatic or labeled with a one if the atom is aromatic. Other atomic features and all bond features are not shown for visual clarity. \textbf{(B)} Data augmentation strategy for polyGNN. Rows of the original training data are transformed by repeat unit addition.}
	\label{fig:Figure_3}
\end{figure}

\subsection{Fingerprinting graphs}\label{sec:graph_fp}
The node features used in this work are element type, node degree, implicit valence, formal charge, number of radical electrons, hybridization, aromaticity (i.e., whether or not a given node is part of an aromatic ring), and number of hydrogen atoms. The edge features are bond type, conjugation (i.e., whether or not a given edge is part of a conjugated system), and ring member (i.e., whether or not a given edge is part of a ring).

\subsection{Neural message passing}\label{sec:passing}
In GNNs, ``messages'' between neighboring atoms in a graph are iteratively passed along chemical bonds. After each iteration, every atom fingerprint is updated using the messages. In this way, atoms learn about their local neighborhood over time. By fitting parameters (e.g., weights and biases) in the model, the information contained in each message is optimized for the task at hand. This process is captured by three general but abstract equations presented in Section S3. In this section, for concreteness, we will demonstrate message passing using a highly simplified example. 

First, consider the graph of infinite polyethylene glycol (PEG), shown in Fig. 1. We restrict our initial atom features to the element type and our initial bond features to the bond type. Thus, all edge fingerprints on the PEG graph are set to [1, 0, 0, 0] (indicating the presence of single bonds and no double, triple, or aromatic bonds). The two carbon atoms in PEG are initialized with a fingerprint of [1, 0] (indicating the presence of C atoms and not O atoms). We index these two nodes 0 and 1. The oxygen atom, with index 2, in PEG is initialized with a fingerprint of [0, 1]. Now, we compute messages $\boldsymbol m_{i,j}$ between all pairs of chemically bonded atoms using the functional form

\begin{equation*}
    \boldsymbol m_{i,j} = ReLU 
    \mleft( 
    W_{\phi} \times
    \mleft[
    \boldsymbol x_{i}^{(0)}, \boldsymbol x_{j}^{(0)}, 
    \boldsymbol e_{i,j}
    \mright]^T
    \mright)
\end{equation*}

where $i,j$ are atom indices, $\boldsymbol x_{i}^{(0)}$ is an initial atom fingerprint, and $\boldsymbol e_{i,j}$ is a bond fingerprint. Note that, for simplicity, we ignore bias terms and use the Rectified Linear Unit (ReLU) activation in this example. $W_{\phi}$ is a matrix of parameters. Before training, the parameters are randomly initialized. During training, the parameters are iteratively updated (i.e., learned) using some flavor of stochastic gradient descent. In this example, our choice of initial parameters will be guided by mathematical convenience, and we do not consider subsequent weight updates. Choosing
\begin{gather*}
    W_{\phi} = 
        \begin{bmatrix}
            1 & 0 & 1 & 0 & 1 & 0 & 0 & 0 \\
            1 & 0 & 1 & 0 & 1 & 0 & 0 & 0 \\
            1 & 0 & 1 & 0 & 1 & 0 & 0 & 0 \\
            1 & 0 & 1 & 0 & 1 & 0 & 0 & 0 \\
            1 & 0 & 1 & 0 & 1 & 0 & 0 & 0 \\
            1 & 0 & 1 & 0 & 1 & 0 & 0 & 0 \\
            1 & 0 & 1 & 0 & 1 & 0 & 0 & 0 \\
            1 & 0 & 1 & 0 & 1 & 0 & 0 & 0 \\
        \end{bmatrix}
    \text{,}
\end{gather*}

gives us

\begin{gather*}
    \begin{split}
        \boldsymbol m_{0,1} = \boldsymbol m_{1,0} = [3,3,3,3,3,3,3,3]        
    \end{split} \\
    \begin{split}
        \boldsymbol m_{0,2} = \boldsymbol m_{1,2} = [2,2,2,2,2,2,2,2]         
    \end{split} \\
    \begin{split}
        \boldsymbol m_{2,0} = \boldsymbol m_{2,1} = [2,2,2,2,2,2,2,2]
    \end{split}
    \text{.}
\end{gather*}

Now, these messages can be used to update the fingerprint of each atom using the functional form

\begin{gather*}
    \begin{split}
        \boldsymbol x_i^{(1)} = ReLU 
        \mleft( 
            W_{\chi} \times
            \mleft[
            \boldsymbol x_{i}^{(0)}, 
            \boldsymbol \sum_j \boldsymbol m_{i,j}
            \mright]^T
        \mright)
    \end{split}
\end{gather*}

where $W_{\chi}$ is a matrix of parameters, and $j$ takes on values corresponding to atoms that share a chemical bond with atom $i$. After we conveniently initialize $W_{\chi}$ to a $2 \times 10$ all-ones matrix, we have
\begin{gather*}
    \begin{split}
        \boldsymbol x_0^{(1)} = [41, 41]        
    \end{split} \\
    \begin{split}
        \boldsymbol x_1^{(1)} = [41, 41]        
    \end{split} \\
    \begin{split}
        \boldsymbol x_2^{(1)} = [33, 33]        
    \end{split}
    \text{.}
\end{gather*}

So, by exchanging messages with neighbors, the fingerprint of each carbon atom in PEG was updated from [1, 0] to [41, 41] and the fingerprint of each oxygen atom was updated from [0, 1] to [33, 33]. The effect of message passing is clear. Initially, the oxygen atom was not aware of neighboring carbon atoms (that is, $\boldsymbol x_{2,2}^{(0)}=0$, where $\boldsymbol x_{i,l}$ is the $l^{th}$ dimension of $\boldsymbol x_{i}$). However, after passing one round of messages, the oxygen atom becomes aware of its carbonaceous neighbors (i.e., $\boldsymbol x_{2,2}^{(1)} \neq 0$). Likewise, the carbon atoms become aware of their neighboring oxygen atom over time.

\section{Results and Discussion}\label{section:results}
\subsection{Benchmarking speed}\label{sec:speed}
polyGNN was developed with a primary objective in mind: to increase the rate at which large libraries of polymers may be screened. We quantified this rate by measuring the time needed to fingerprint a data set of 13,338 known polymers on a variety of different capacities and hardware. Capacity, as used in this work, is a hyperparameter that specifies both the number of message passing steps and the depth of each multilayer perceptron (MLP) in the network. 

The timings to compute 13,388 polymer fingerprints with a randomly initialized polyGNN model are given in Fig. 4. A shallow polyGNN (with a capacity of two) fingerprints the set of polymers in 32 seconds (2.4 ms per polymer) on one CPU or 30 seconds (2.2 ms per polymer) on one GPU. Meanwhile, a deep polyGNN (with a capacity of 12) takes 57 seconds (4.3 ms per polymer) to compute the fingerprint set on one CPU or 32 seconds on one GPU. For each of the above, the time spent on the Encoder was fixed at 26 seconds. The remaining time was spent on the Message Passing Block which, unlike the Encoder, can run on CPUs or graphics processing units (GPUs). 

\begin{figure}[ht]
    \centering
	\includegraphics[width=0.99\textwidth]{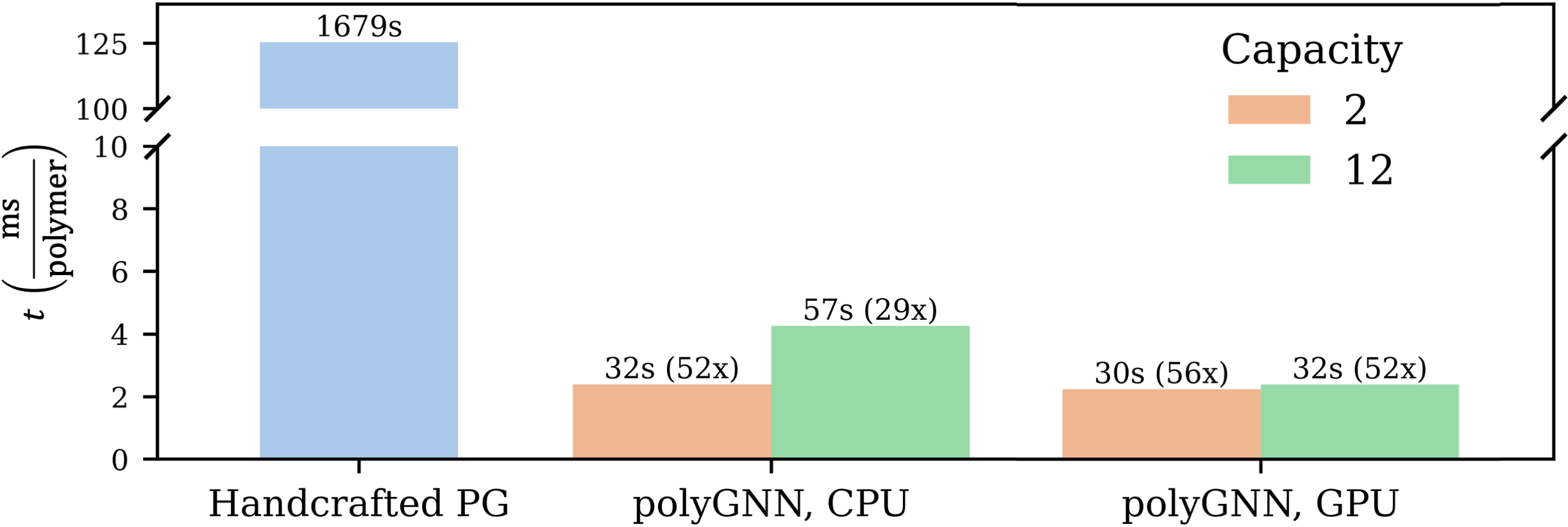}
	\caption{Fingerprint time as a function of method, capacity, and hardware. Fingerprint time $t$, measured in milliseconds per polymer, is plotted on the $y$-axis. $t$ was computed using a diverse set of 13,388 polymers. Above each bar is the total time (in plain text) in seconds taken to compute fingerprints for the entire set as well as the speed up (in parentheses) relative to the handcrafted PG method. $t$ varies depending on the fingerprint method, the hardware, and the model capacity. Method and hardware are labelled on the $x$-axis; CPU and GPU refer to one Intel$^\text{{\tiny{\textregistered}}}$ Xeon$^\text{{\tiny{\textregistered}}}$ Gold 6140 CPU core and to one 32 GB Nvidia$^\text{{\tiny{\textregistered}}}$ Tesla$^\text{{\tiny{\textregistered}}}$ V100-PCIE GPU, respectively. Capacity is denoted by bar color.}
	\label{fig:speed}
\end{figure}

By extrapolation, this means that fingerprinting a library of 1 billion polymers would take 26 days in the best case (shallow polyGNN run on a GPU) and 47 days in the worst case (deep polyGNN run on one CPU). Meanwhile, at a rate of 125.4 ms per polymer, fingerprinting a library of 1 billion polymers would take nearly 4 years on one CPU using the handcrafted PG approach. Of course, the rates for either approach can be further sped up with parallelization and/or increased random access memory.  

\subsection{Benchmarking accuracy}
Here we evaluate the predictive accuracy of polyGNN models on 34 of the 36 properties in our data set; dielectric constant at $10^7$ and $10^9$ Hz ($\epsilon_\text{7}$ and $\epsilon_\text{9}$) were excluded because our corpus contains fewer than 50 data points for these properties. Data for the remaining properties was randomly cut into a training and a test set in a $4:1$ ratio. Three such random cuts were performed per property, so that statistics (e.g., standard deviation) of model performance could be computed.

Kuenneth et al. \cite{Kuenneth2021PolymerLearning} showed that multitask learning significantly improves the accuracy of polymer property prediction, relative to single task learning. Thus, we train single task (ST) and multitask (MT) polyGNNs and compare both on the same data. As a benchmark, we also train both ST and MT ``PG-MLPs'' (i.e., MLPs that use the handcrafted PG fingerprint as input; see Section S4 for details on this architecture). A detailed discussion of our training procedure can be found in Section S5. The root-mean-squared-error (RMSE) and R$^2$ values of polyGNN and PG-MLP are compared in Tables \ref{table:rmse_values} and S1.

We note several observations from these results. First, our data augmentation strategy plays a critical role in teaching polyGNN models invariance to addition and subtraction (see Table S2). Second, we find that MT learning is an important component of our approach, especially in low data situations. As shown in Table S1, polyGNNs that do not use MT learning exhibit erroneous predictions (i.e., negative R$^2$ value) for five properties\textemdash $E_\text{i}$, $\epsilon_\text{1.78}$, $\epsilon_\text{2}$, $\epsilon_\text{5}$, $\epsilon_\text{6}$\textemdash each with 158 or fewer data points. In contrast, with MT learning, polyGNNs exhibit positive R$^2$ for each of the 34 properties studied.

\begin{table*}
\centering
\scalebox{0.99}{
\begin{threeparttable}
\begin{tabular}{l|rrrr}
  \toprule
  Property & MT polyGNN & MT PG-MLP & ST polyGNN & ST PG-MLP\\
  \midrule
  $\lambda$* & \textbf{0.0547} $\pm$ 0.0103 & 0.0630 $\pm$ 0.0082 & 0.0580 $\pm$ 0.0096 & 0.0663 $\pm$ 0.0201\\
  $T_\text{m}$ & \textbf{45.0} $\pm$ 1.8 & \textbf{47.2} $\pm$ 2.2 & 55.3 $\pm$ 2.8 & 53.1 $\pm$ 1.3\\
  $T_\text{d}$ & \textbf{58.7} $\pm$ 3.3 & \textbf{59.3} $\pm$ 2.0 & 67.7 $\pm$ 3.2 & 71.9 $\pm$ 6.9\\
  $T_\text{g}$ & \textbf{31.7} $\pm$ 1.5 & \textbf{34.0} $\pm$ 0.9 & \textbf{36.6} $\pm$ 1.0 & \textbf{35.5} $\pm$ 1.6\\
  \midrule
  $E_\text{at}$* & \textbf{0.114} $\pm$ 0.071 & 0.284 $\pm$ 0.089 & \textbf{0.0913} $\pm$ 0.0224 & 0.155 $\pm$ 0.040\\
  $c_\text{p}$* & \textbf{0.172} $\pm$ 0.033 & 0.223 $\pm$ 0.085 & \textbf{0.171} $\pm$ 0.019 & \textbf{0.161} $\pm$ 0.030\\
  $O_\text{i}$* & \textbf{8.99} $\pm$ 1.01 & 9.77 $\pm$ 1.57 & \textbf{8.79} $\pm$ 0.46 & \textbf{8.63} $\pm$ 0.47\\
  $X_\text{e}$* & 15.0 $\pm$ 3.7 & \textbf{13.1} $\pm$ 4.6 & 15.8 $\pm$ 3.9 & 17.1 $\pm$ 5.1\\
  $V_\text{ff}$* & 0.0380 $\pm$ 0.0191 & 0.0423 $\pm$ 0.0216 & \textbf{0.0330} $\pm$ 0.0182 & 0.0373 $\pm$ 0.0215\\
  $X_\text{c}$ & \textbf{16.6} $\pm$ 1.3 & \textbf{17.4} $\pm$ 2.5 & 18.6 $\pm$ 1.9 & 19.1 $\pm$ 2.2\\
  $\rho$ & \textbf{0.0640} $\pm$ 0.0053 & 0.0937 $\pm$ 0.0025 & \textbf{0.0627} $\pm$ 0.0015 & 0.385 $\pm$ 0.264\\
  \midrule
  $E_\text{a}$* & 0.380 $\pm$ 0.034 & 0.483 $\pm$ 0.148 & \textbf{0.341} $\pm$ 0.055 & \textbf{0.357} $\pm$ 0.107\\
  $E_\text{i}$* & \textbf{0.540} $\pm$ 0.170 & 0.678 $\pm$ 0.231 & 59.9 $\pm$ 102.5 & 0.676 $\pm$ 0.139\\
  $E_\text{gb}$* & \textbf{0.468} $\pm$ 0.066 & \textbf{0.535} $\pm$ 0.123 & 0.716 $\pm$ 0.164 & 0.737 $\pm$ 0.058\\
  $E_\text{gc}$ & \textbf{0.445} $\pm$ 0.018 & \textbf{0.491} $\pm$ 0.033 & \textbf{0.442} $\pm$ 0.020 & \textbf{0.494} $\pm$ 0.026\\
  \midrule
  $\epsilon_\text{0}$* & \textbf{0.285} $\pm$ 0.101 & \textbf{0.284} $\pm$ 0.061 & 0.362 $\pm$ 0.086 & \textbf{0.252} $\pm$ 0.014\\
  $\epsilon_\text{1.78}$* & 0.427 $\pm$ 0.042 & \textbf{0.328} $\pm$ 0.067 & 1.34 $\pm$ 0.30 & 0.988 $\pm$ 0.517\\
  $\epsilon_\text{2}$* & 0.478 $\pm$ 0.228 & \textbf{0.376} $\pm$ 0.257 & 2.67 $\pm$ 2.78 & 0.937 $\pm$ 0.201\\
  $\epsilon_\text{3}$* & \textbf{0.621} $\pm$ 0.250 & 0.806 $\pm$ 0.338 & 1.39 $\pm$ 0.21 & 1.42 $\pm$ 0.22\\
  $\epsilon_\text{4}$* & \textbf{0.284} $\pm$ 0.018 & \textbf{0.252} $\pm$ 0.030 & 0.650 $\pm$ 0.108 & 0.602 $\pm$ 0.175\\
  $\epsilon_\text{5}$* & \textbf{0.212} $\pm$ 0.023 & \textbf{0.243} $\pm$ 0.011 & 0.479 $\pm$ 0.266 & 0.658 $\pm$ 0.358\\
  $\epsilon_\text{6}$* & 0.323 $\pm$ 0.075 & \textbf{0.274} $\pm$ 0.034 & 0.676 $\pm$ 0.315 & 0.487 $\pm$ 0.214\\
  $\epsilon_\text{15}$ & \textbf{0.125} $\pm$ 0.015 & 0.145 $\pm$ 0.019 & 0.144 $\pm$ 0.021 & 0.171 $\pm$ 0.027\\
  $n_\text{c}$* & \textbf{0.0507} $\pm$ 0.0186 & 0.0733 $\pm$ 0.0191 & 0.0933 $\pm$ 0.0304 & 0.0957 $\pm$ 0.0251\\
  $n_\text{e}$ & \textbf{0.0413} $\pm$ 0.0023 & \textbf{0.0437} $\pm$ 0.0090 & 0.0540 $\pm$ 0.0087 & 0.0760 $\pm$ 0.0262\\
  \midrule
  $Y$ & \textbf{0.827} $\pm$ 0.099 & \textbf{0.760} $\pm$ 0.169 & 0.877 $\pm$ 0.074 & 0.860 $\pm$ 0.196\\
  $\sigma_\text{ts}$ & \textbf{23.3} $\pm$ 5.5 & \textbf{22.2} $\pm$ 3.9 & 28.1 $\pm$ 4.6 & 25.8 $\pm$ 3.9\\
  \midrule
  $\delta_\text{s}$* & \textbf{1.15} $\pm$ 0.11 & 2.11 $\pm$ 0.10 & 1.65 $\pm$ 0.33 & 1.36 $\pm$ 0.09\\
  $\mu_\text{He}$* & \textbf{0.133} $\pm$ 0.017 & \textbf{0.111} $\pm$ 0.014 & 0.265 $\pm$ 0.065 & 0.246 $\pm$ 0.011\\
  $\mu_{\text{H}_2}$* & \textbf{0.127} $\pm$ 0.006 & \textbf{0.104} $\pm$ 0.011 & 0.287 $\pm$ 0.013 & 0.367 $\pm$ 0.034\\
  $\mu_{\text{CO}_2}$ & \textbf{0.166} $\pm$ 0.015 & \textbf{0.161} $\pm$ 0.019 & 0.430 $\pm$ 0.025 & 0.525 $\pm$ 0.212\\
  $\mu_{\text{CH}_4}$ & \textbf{0.132} $\pm$ 0.024 & \textbf{0.113} $\pm$ 0.023 & 0.366 $\pm$ 0.030 & 0.397 $\pm$ 0.006\\
  $\mu_{\text{N}_2}$ & \textbf{0.124} $\pm$ 0.011 & \textbf{0.109} $\pm$ 0.018 & 0.410 $\pm$ 0.104 & 0.397 $\pm$ 0.038\\
  $\mu_{\text{O}_2}$ & \textbf{0.139} $\pm$ 0.014 & \textbf{0.114} $\pm$ 0.004 & 0.399 $\pm$ 0.062 & 1.83 $\pm$ 2.46\\
  \bottomrule
\end{tabular}
\begin{tablenotes}\footnotesize
\item[$\dagger$] Starred properties contain 300 or fewer data points. Models with the best, or comparable with the best, average RMSE are bolded. The unit of each RMSE value matches those listed in Fig. \ref{fig:dataset}a; for example, the RMSE of the MT polyGNN approach on $T_{\text{g}}$ is 31.7 $\pm$ 1.5 K.
\end{tablenotes}
\end{threeparttable}
}
\caption{\textbf{Average RMSE plus/minus one standard deviation on unseen test data.}\textsuperscript{$\dagger$}}
\label{table:rmse_values}
\end{table*}

Third, we find that polyGNNs tend to exbihit better or comparable accuracy than PG-MLPs, especially when the number of training data points is greater than 300. For the 14 properties containing more than 300 data points, each MT polyGNN model is either more accurate than or comparably accurate to its corresponding MT PG-MLP model (we define two models as having comparable accuracy for a property if the difference in average RMSE of their predictions is within 5\% of that property's standard deviation $\sigma$, see Table S3 for a complete list of $\sigma$ values). However, for the 20 properties containing 300 data points or less, the situation becomes more complex. MT polyGNN models still perform well relative to the MT PG-MLP benchmark, but not for every property. MT polyGNN models are more or comparably accurate for 16 properties, but are notably less accurate on four properties (experimental crystallization tendency $X_\text{e}, \epsilon_\text{1.78}, \epsilon_\text{2}, \epsilon_\text{6}$).

The relatively low performance on these four properties could be explained by the fact that the polyGNN models trained here struggle to learn the block- or chain-level features (which typically consist of 4+ atoms) present in the handcrafted PG fingerprint. In principle, increasing the number of message passing steps\textemdash so as to capture larger length scale features\textemdash should mitigate this challenge. In practice, however, we observe a threshold number of message passing steps. Above three message passing steps, model generalization only worsens\textemdash regardless of the property of interest. This empirical observation has been reported by others and is due to a collapse in which the learned fingerprints of all polymers, even chemically distinct ones, converge \cite{Godwin2021SimpleBeyond, Chen2019MeasuringView}. However, as evidenced by the impressive performance of the MT polyGNN models on a vast majority of properties, the inability to learn block- or chain-level features features is often ameliorated by the ability to learn lower-level features that go beyond those currently present in the handcrafted PG fingerprint. Still, the development of techniques that encourage GNNs to surpass the message passing threshold is a critical next step. We leave this task for future work.

\section{Summary and Outlook}
In summary, we have produced polyGNN\textemdash the first-ever protocol that integrates polymer feature learning from SMILES strings and other relevant features, invariant transformations, data augmentation and multitask learning. Through careful comparison, we show that our protocol culminates in ultrafast polymer fingerprinting and accurate property prediction over the most comprehensive array of chemistries and properties studied to date. The gains in speed are essential when screening large candidate sets (e.g., millions or billions of polymers) and/or when computational resources are limited. Our approach is especially accurate when the data set size is moderate to large. Even with data sets containing less than 300 points, our approach is at least competitive with presently adopted methods in a majority of cases.

Looking ahead, though polyGNNs perform remarkably well in the experiments tried here, handcrafted polymer fingerprints have advantages. In tasks where chain- or block-level features are essential, handcrafted fingerprinting approaches may yield the best model accuracy. Advances in the optimization of graph neural networks are needed to make the accuracy of polyGNNs competitive in these tasks. Finally, a handcrafted feature is, by definition, interpretable. In contrast, the features learned by the polyGNNs presented here are not interpretable. Following the work of others \cite{Velickovic2017GraphNetworks}, future polyGNN architectures may incorporate attention mechanisms for partial interpretability. However, the interpretability of polyGNN features at the level of handcrafted features will require further innovation. Despite these shortcomings, we anticipate that the adoption of polyGNNs and related approaches will increase as they unlock the ability to screen truly massive polymer libraries at scale.

\section{Public Use}
The sources of data used in this work and the availability of each source is reported in the paper. The code used to train our polyGNN models is available at \texttt{github.com/Ramprasad-Group/polygnn} for academic use.

\begin{acknowledgement}



This work was financially supported by the Office of Naval Research through a Multi-University Research Initiative (MURI) grant (N00014-17-1-2656), the Center for Understanding and Control of Acid Gas Induced Evolution of Materials for Energy (UNCAGE ME, an Energy Frontier Research Center) funded by the U.S. Department of Energy (DOE) under Award \# DE-SC0012577, and by the National Science Foundation under grant 1941029. C.K. thanks the Alexander von Humboldt Foundation for financial support. R.G. is the main architect of the machine learning models and wrote this paper. C.K. and A.T. supported in the development and debugging of the machine learning models. The work was conceived and guided by R.R. All authors discussed results and commented on the manuscript.

\end{acknowledgement}

\bibliography{references}

\section*{Supporting Information}
\renewcommand{\thesection}{S\arabic{section}}
\renewcommand\thefigure{S\arabic{figure}}
\renewcommand\thetable{S\arabic{table}}
\setcounter{section}{0}
\setcounter{figure}{0}
\setcounter{table}{0}

\section{Data breakdown}
Our data set includes the permeability $\mu$ of six gases $g \in$ \{\ch{He}, \ch{H2}, \ch{CO2}, \ch{O_2}, \ch{N_2}, \ch{CH_4}\}. The number of data points per gas is: 281 for \ch{He}, 288 for \ch{H2}, 342 for \ch{CO2}, 380 for \ch{CH4}, 431 for \ch{N2}, and 436 for \ch{O2}. Our experimental dielectric constant $\epsilon_{\text{f}}$ data contains measurements at nine frequencies $f \in \{1.78, 2, 3, 4, 5, 6, 7, 9, 15\}$ in $\text{log}_{10}$Hz. The number of data points per frequency is: 51 for $10^{1.78}$ Hz, 77 for $10^{2}$ Hz, 172 for $10^{3}$ Hz, 124 for $10^{4}$ Hz, 66 for $10^{5}$ Hz, 158 for $10^{6}$ Hz, 20 for $10^{7}$ Hz, 12 for $10^{9}$ Hz, 507 for $10^{15}$ Hz.  

\section{polyGNN2 Encoder}
\begin{figure}[h]
    \centering
    \noindent\includegraphics[width=0.8\textwidth]{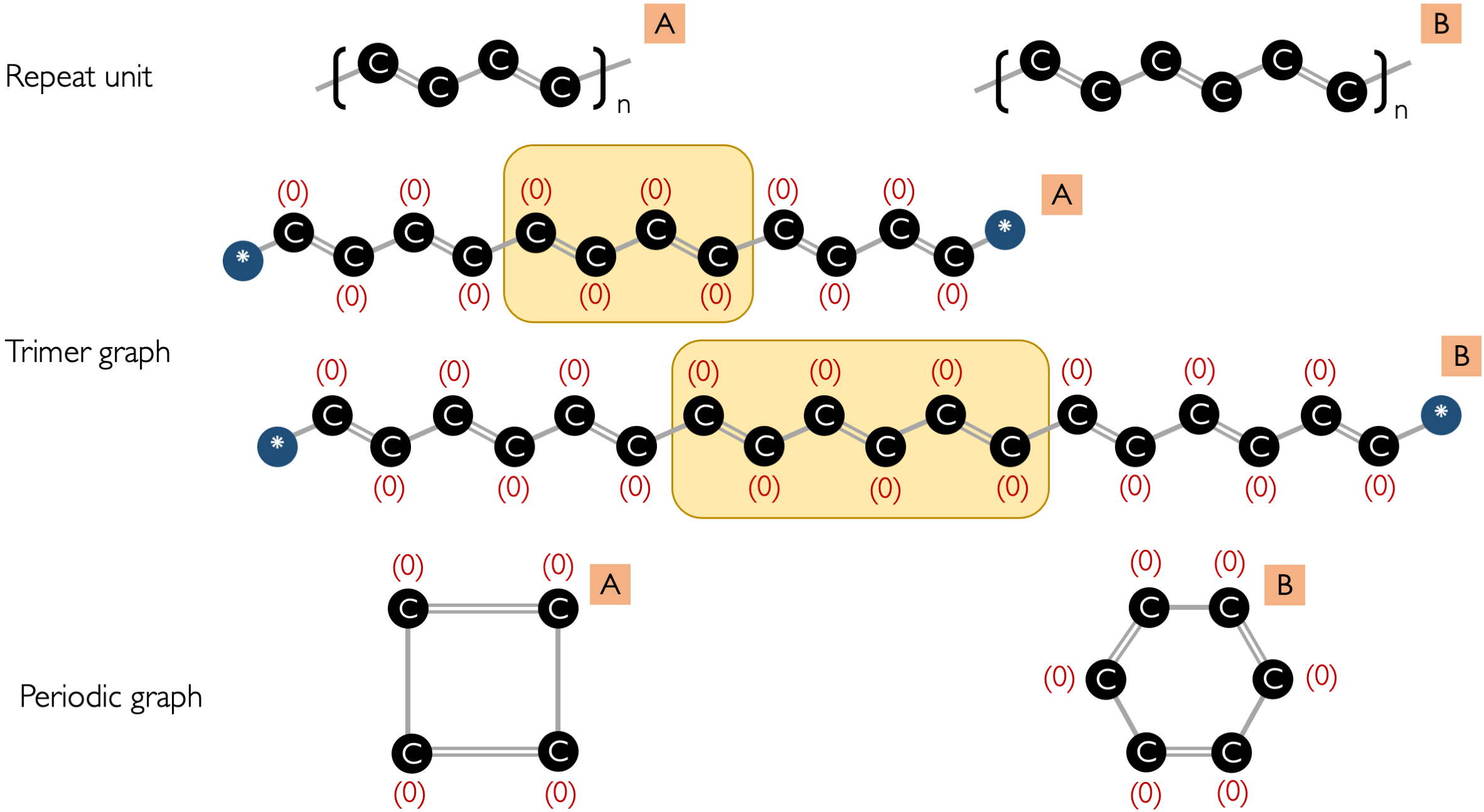}
    \caption{Two equivalent repeat units (``A'' and ``B'', where ``A'' and ``B'' refer to four and six atom repeat units, respectively) of polyacetylene and their corresponding trimer and periodic graphs. Each repeat unit is converted to a trimer graph. Each node (i.e., heavy atom) in the trimer graph is featurized. Each atom is labeled with a zero if the atom is aliphatic or is labeled with a one if the atom is aromatic. Other atomic features and all bond features are not shown for visual clarity. The atoms, bonds, and features at the center of the trimer graph (shaded in yellow) are used to form the periodic graph.}
    \label{fig:polyGNN2}
\end{figure}

The input to the polyGNN2 Encoder is a repeat unit. From this repeating unit, the trimer graph (shown in Figure \ref{fig:polyGNN2}) is created. Then, the trimer graph is featurized. The atoms, bonds, and features corresponding to the central repeat unit of the trimer graph are used to create the periodic graph. Thus, the initial fingerprint of the periodic graph is always invariant to addition and subtraction, as shown for the example of polyacetylene in Figure \ref{fig:polyGNN2}. However, the polyGNN2 Encoder is slower than the polyGNN Encoder. While the polyGNN Encoder takes 26 seconds to featurize the graphs of $13\,338$ polymers, the polyGNN2 Encoder takes 37 seconds to featurize this set of polymer graphs.

\section{polyGNN architecture}
polyGNNs contain three modules: the Encoder, Message Passing Block, and the Estimator. The inputs to polyGNN are a polymer repeat unit and a property of interest (or, equivalently, the property's associated selector vector). The two outputs of a polyGNN model are the repeat unit's fingerprint and the value of the property of interest. 

In the Encoder, the repeat unit is first converted to a periodic graph, with each atom as a node and each bond as an edge. Then, each node and edge in the graph are given an initial fingerprint. After the graph elements have been assigned their initial features, the graph is passed to the Message Passing Block. Here, ``messages'' between neighboring atoms are iteratively passed along chemical bonds. After each iteration, every node fingerprint is updated using the messages, while each bond fingerprint remains the same. The message passed from atom $j$ to atom $i$ at time step $k$ is calculated according to Eq. \ref{eq:message}.

\begin{equation}\label{eq:message}
    \boldsymbol m^{(k)}_{i,j} = \phi^{(k)} 
    \mleft( 
    \boldsymbol x_{i}^{(k)}, \boldsymbol x_{j}^{(k)}, 
    \boldsymbol e_{i,j}
    \mright)
\end{equation}

where each $\phi^{(k)}$ is a parameterized function, $\boldsymbol x_{i}^{(k)}$ and $\boldsymbol x_{j}^{(k)}$ are the encodings of neighboring atoms after time step $k$, and $\boldsymbol e_{i,j}$ is the fingerprint of the bond that joins atoms $i,j$. $\boldsymbol m^{(k)}_{i,j} = 0$ if $i,j$ do not share a chemical bond. After initialization, each node receives messages from all of its neighbors. These messages are aggregated by some permutation-invariant function $\mathcal{I}$ (e.g., \texttt{sum}, \texttt{mean}, \texttt{max}). We use the \texttt{sum} in this work. The aggregated message, along with the current node encoding, is used to \emph{update} the node encoding. The node update process is defined in Equation \ref{eq:update}.

\begin{equation}\label{eq:update}
    \boldsymbol x_i^{(k)} = \chi^{(k)} \mleft( 
    \boldsymbol x_{i}^{(k-1)}, 
    \mathcal{I}
    \mleft( 
    \{\boldsymbol m_{i,j} \forall j \in [\![1,N_p]\!]\}
    \mright) 
    \mright) +
    x_{i}^{(k-2)}
\end{equation}

where each $\chi^{(k)}$ is a parameterized function, $p$ is a polymer, $[\![1,N_p]\!]$ is the set of integers between 1 and $N_p$, $N_p$ is the number of atoms in the repeat unit of $p$, and $\boldsymbol x_i^{(k)} = 0, \forall k < 0$. Messages are passed for $\tau$ time steps, where $\tau$ is also the capacity in this work. The fingerprint of the entire polymer, $\boldsymbol x_p$, is calculated by the graph aggregation function $\mathcal{A}_g$, as shown in Eq. \ref{eq:mpnn_fp}.

\begin{equation}\label{eq:mpnn_fp}
    \boldsymbol x_p = \mathcal{A}_g(x_{i}^{(\tau)}, x_{i}^{(0)}) = \dfrac{1}{N_p}\sum_{i=1}^{N_p} \boldsymbol x_{i}^{(\tau)} + \boldsymbol x_{i}^{(0)}
\end{equation}

Finally, $\boldsymbol x_p$ and the selector $\boldsymbol s$ can be passed to the Estimator. Here, these inputs are mapped to some polymer property prediction, $y_p$, via a parameterized function $\psi$. We implement $\psi$ as a multilayer perceptron.

\begin{equation}
    y_p = \psi (\boldsymbol x_p, \boldsymbol s)
\end{equation}

During training, the parameters of all $\phi^{(k)}, \chi^{(k)}, \psi$ are learned \emph{simultaneously}. As shown in Eq. \ref{eq:update}, our update step leverages skip connections, which have been shown to improve the optimization of shallow layers in deep neural networks \cite{He2016DeepRecognition}.  

\section{Handcrafted PG models}\label{sec:pg_architecture}
The handcrafted PG models are made up of five MLP submodels (see Section S4), trained using five-fold cross-validation. The input to each MLP is the handcrafted PG fingerprint of a given polymer repeat unit and a property selector and the output is the predicted property value.

\section{Training procedure}
\begin{figure}[h]
    \centering
    \noindent\includegraphics[width=0.8\textwidth]{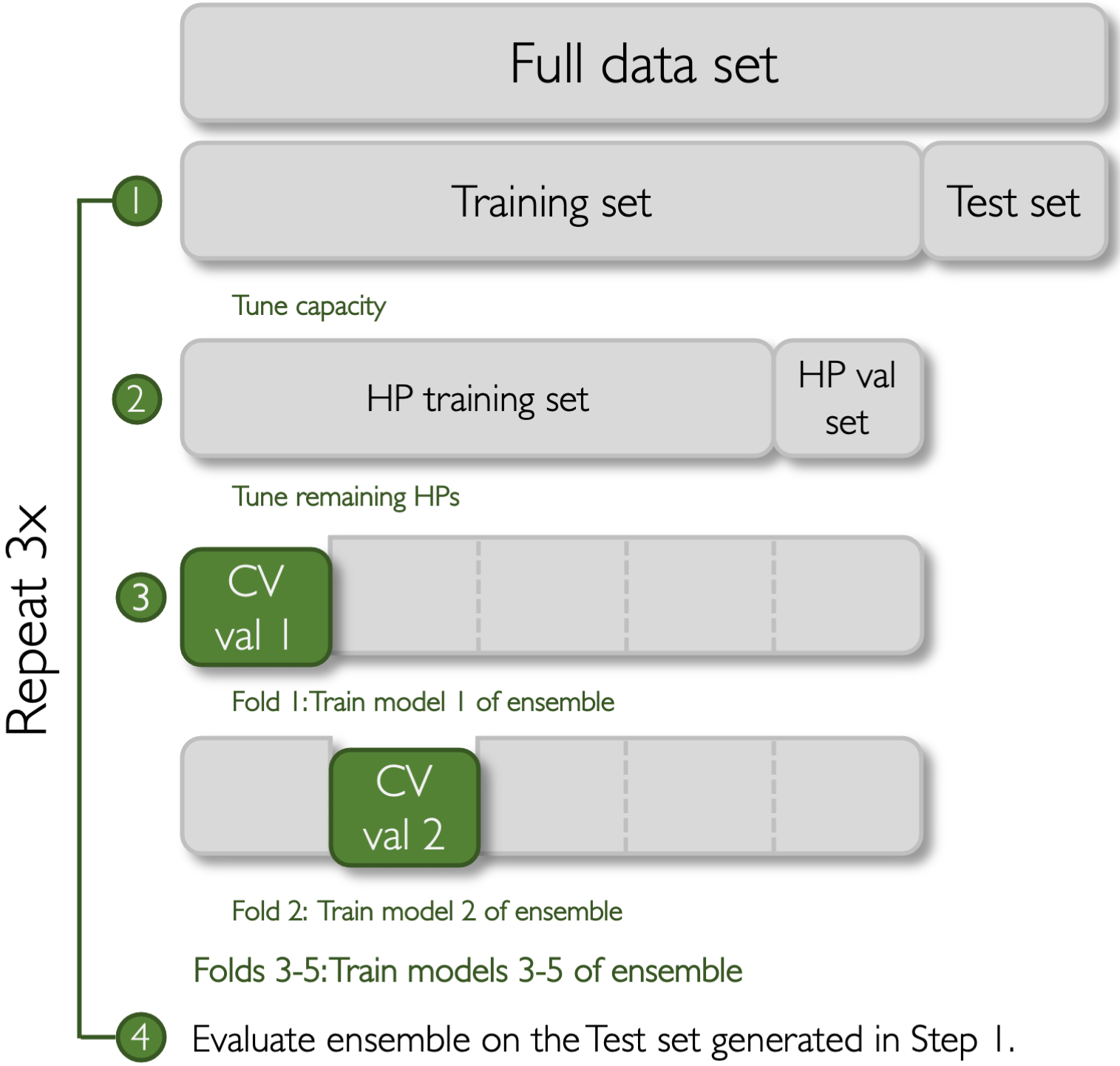}
    \caption{The various steps in our training and evaluation protocol. This protocol involved three runs, with three rounds of data splitting per run. All splits are random. ``HP'' stands for hyperparameter, ``val'' stands for validation, and ``CV'' stands for cross-validation.}
    \label{fig:Data_split}
\end{figure}

Each of the models discussed in the main text are ensemble models, composed of several submodels. The output of the ensemble is computed by a simple average of each submodel's output. For multitask ensembles, data for all properties within a group were combined, target values were scaled, and selectors were assigned as described in Section 2.1 of the main text. For single-task ensembles, properties were not combined into groups, MinMax scaling was not performed, and selectors were set to empty vectors. Next, to train and evaluate the model, the data was split according to the schematic in Figure \ref{fig:Data_split} described below.

First, the entire data set for each property was randomly cut into 80\% training, 20\% test splits three times. All subsequent steps were performed for each training-test set pair. Using the \texttt{NNDebugger} package \cite{35339}, the optimal capacity was found by attempting to overfit the entire training data set. The data set was considered overfit if the R$^2$ value was greater than 0.97. If the data was not overfit, then the capacity corresponding to the highest R$^2$ value was used. The capacity range considered was between two and fourteen. The training data set was then divided into an 80\% hyperparameter (HP) training set and a 20\% HP validation set. The remaining HPs (batch size, learning rate, dropout percentage) were optimized using the package \texttt{scikit-optimize}. The set of HPs corresponding to the lowest RMSE on the HP validation set was considered optimal.

Finally, the training data set was split into five folds using cross-validation (CV), producing one CV train data set and one CV validation data set per fold. For each fold, the model's HPs were fixed as the optimal HPs and the model's learnable parameters were fit to the CV train data set for 1000 epochs. At the end of 1000 epochs, the model parameters corresponding to the epoch with the lowest RMSE in the CV validation data set were chosen. After all five models were trained on their respective CV splits, the models were placed in an ensemble. The ensemble was used to make predictions of the test set, so far completely unseen by the ensemble during HP optimization or model training with CV.

All neural network architectures used dropout layers, fully connected layers, and Leaky ReLU activations (with a negative slope equal to $0.01$). All architectures were created using \texttt{PyTorch} and/or \texttt{PyTorch Geometric}. The weights of all models were optimized using the Adam optimizer and the mean squared error loss function. All weights were initialized according to a Xavier uniform distribution \cite{pmlr-v9-glorot10a} with a gain of one. All biases were initialized using the default \texttt{PyTorch} setting.

\section{Extended Results}
We computed both the root-mean-squared-error (RMSE) and Pearson correlation coefficient (R$^2$) of each trained ensemble on unseen data for each property. The RMSE values are tabulated in the main body and the R$^2$ values are tabulated in Table \ref{table:R2_values}.

\begin{table}[h]
\centering
\begin{tabular}{l|rrrr}
  \toprule
  Property & MT polyGNN & MT handcrafted PG & ST polyGNN & ST handcrafted PG\\
  \midrule
  $\lambda$* & 0.106 $\pm$ 0.245 & -0.177 $\pm$ 0.205 & 0.002 $\pm$ 0.209 & -0.367 $\pm$ 0.678\\
  $T_\text{m}$ & 0.840 $\pm$ 0.019 & 0.824 $\pm$ 0.023 & 0.757 $\pm$ 0.032 & 0.777 $\pm$ 0.018\\
  $T_\text{d}$ & 0.746 $\pm$ 0.019 & 0.741 $\pm$ 0.021 & 0.662 $\pm$ 0.019 & 0.618 $\pm$ 0.059\\
  $T_\text{g}$ & 0.913 $\pm$ 0.004 & 0.901 $\pm$ 0.004 & 0.885 $\pm$ 0.011 & 0.892 $\pm$ 0.004\\
  \midrule
  $E_\text{at}$* & 0.933 $\pm$ 0.057 & 0.614 $\pm$ 0.107 & 0.959 $\pm$ 0.011 & 0.879 $\pm$ 0.058\\
  $c_\text{p}$* & 0.729 $\pm$ 0.135 & 0.541 $\pm$ 0.280 & 0.734 $\pm$ 0.125 & 0.757 $\pm$ 0.142\\
  $O_\text{i}$* & 0.561 $\pm$ 0.092 & 0.475 $\pm$ 0.166 & 0.581 $\pm$ 0.057 & 0.596 $\pm$ 0.058\\
  $X_\text{e}$* & 0.227 $\pm$ 0.112 & 0.432 $\pm$ 0.137 & 0.143 $\pm$ 0.133 & 0.018 $\pm$ 0.210\\
  $V_\text{ff}$* & 0.380 $\pm$ 0.319 & 0.259 $\pm$ 0.355 & 0.534 $\pm$ 0.265 & 0.417 $\pm$ 0.351\\
  $X_\text{c}$ & 0.519 $\pm$ 0.025 & 0.479 $\pm$ 0.078 & 0.397 $\pm$ 0.073 & 0.368 $\pm$ 0.072\\
  $\rho$ & 0.894 $\pm$ 0.022 & 0.777 $\pm$ 0.034 & 0.900 $\pm$ 0.014 & -4.126 $\pm$ 4.386\\
  \midrule 
  $E_\text{a}$* & 0.738 $\pm$ 0.114 & 0.613 $\pm$ 0.105 & 0.781 $\pm$ 0.108 & 0.782 $\pm$ 0.079\\
  $E_\text{i}$* & 0.786 $\pm$ 0.156 & 0.659 $\pm$ 0.263 & -6553.467 $\pm$ 11352.055 & 0.681 $\pm$ 0.162\\
  $E_\text{gb}$* & 0.928 $\pm$ 0.035 & 0.908 $\pm$ 0.040 & 0.839 $\pm$ 0.067 & 0.828 $\pm$ 0.046\\
  $E_\text{gc}$ & 0.915 $\pm$ 0.002 & 0.896 $\pm$ 0.014 & 0.916 $\pm$ 0.005 & 0.895 $\pm$ 0.006\\
  \midrule
  $\epsilon_\text{0}$* & 0.557 $\pm$ 0.358 & 0.652 $\pm$ 0.088 & 0.354 $\pm$ 0.404 & 0.683 $\pm$ 0.197\\
  $\epsilon_\text{1.78}$* & 0.842 $\pm$ 0.102 & 0.896 $\pm$ 0.088 & -0.494 $\pm$ 0.851 & 0.224 $\pm$ 0.503\\
  $\epsilon_\text{2}$* & 0.750 $\pm$ 0.252 & 0.807 $\pm$ 0.255 & -12.141 $\pm$ 21.000 & 0.166 $\pm$ 0.302\\
  $\epsilon_\text{3}$* & 0.819 $\pm$ 0.144 & 0.704 $\pm$ 0.199 & 0.203 $\pm$ 0.125 & 0.168 $\pm$ 0.116\\
  $\epsilon_\text{4}$* & 0.861 $\pm$ 0.071 & 0.888 $\pm$ 0.062 & 0.346 $\pm$ 0.178 & 0.441 $\pm$ 0.232\\
  $\epsilon_\text{5}$* & 0.797 $\pm$ 0.088 & 0.720 $\pm$ 0.149 & -0.553 $\pm$ 2.096 & -1.899 $\pm$ 3.879\\
  $\epsilon_\text{6}$* & 0.519 $\pm$ 0.219 & 0.670 $\pm$ 0.078 & -0.913 $\pm$ 1.112 & 0.013 $\pm$ 0.531\\
  $\epsilon_\text{15}$ & 0.860 $\pm$ 0.042 & 0.813 $\pm$ 0.060 & 0.815 $\pm$ 0.066 & 0.742 $\pm$ 0.084\\
  $n_\text{c}$* & 0.874 $\pm$ 0.061 & 0.737 $\pm$ 0.083 & 0.543 $\pm$ 0.307 & 0.532 $\pm$ 0.217\\
  $n_\text{e}$ & 0.850 $\pm$ 0.020 & 0.826 $\pm$ 0.085 & 0.738 $\pm$ 0.083 & 0.437 $\pm$ 0.417\\
  \midrule
  $Y$ & 0.501 $\pm$ 0.107 & 0.587 $\pm$ 0.094 & 0.428 $\pm$ 0.176 & 0.480 $\pm$ 0.073\\
  $\sigma_\text{ts}$ & 0.638 $\pm$ 0.212 & 0.675 $\pm$ 0.155 & 0.499 $\pm$ 0.173 & 0.576 $\pm$ 0.152\\
  \midrule
  $\delta_\text{s}$* & 0.770 $\pm$ 0.074 & 0.235 $\pm$ 0.180 & 0.536 $\pm$ 0.156 & 0.686 $\pm$ 0.045\\
  $\mu_\text{He}$* & 0.969 $\pm$ 0.002 & 0.978 $\pm$ 0.007 & 0.877 $\pm$ 0.037 & 0.891 $\pm$ 0.023\\
  $\mu_{\text{H}_2}$* & 0.983 $\pm$ 0.002 & 0.988 $\pm$ 0.002 & 0.914 $\pm$ 0.018 & 0.857 $\pm$ 0.041\\
  $\mu_{\text{CO}_2}$ & 0.980 $\pm$ 0.006 & 0.981 $\pm$ 0.006 & 0.866 $\pm$ 0.027 & 0.779 $\pm$ 0.174\\
  $\mu_{\text{CH}_4}$ & 0.986 $\pm$ 0.005 & 0.990 $\pm$ 0.004 & 0.897 $\pm$ 0.029 & 0.881 $\pm$ 0.009\\
  $\mu_{\text{N}_2}$ & 0.985 $\pm$ 0.003 & 0.988 $\pm$ 0.003 & 0.833 $\pm$ 0.070 & 0.844 $\pm$ 0.034\\
  $\mu_{\text{O}_2}$ & 0.981 $\pm$ 0.003 & 0.987 $\pm$ 0.002 & 0.845 $\pm$ 0.028 & -6.798 $\pm$ 13.235\\
  \bottomrule
\end{tabular}
\caption{Average R$^2$ plus/minus one standard deviation on unseen test data. Starred properties contain 300 or fewer data points.}
\label{table:R2_values}
\end{table}

As discussed in the main body, we desire that the final output of each polyGNN model is approximately invariant to addition and subtraction. In other words, the variance in predictions between a set of equivalent repeat units should be low. As shown in Table \ref{table:variance}, we find that our proposed data set augmentation does lead to a significant decrease in prediction variance a majority of the time.

\begin{table*}[h!]
\centering
\scalebox{0.95}{
\begin{tabular}{l|rrr}
  \toprule
  Property & $\widehat{var}_\text{no augment}$ & $\widehat{var}_\text{augment}$ & $\dfrac{\widehat{var}_\text{no augment}}{\widehat{var}_\text{augment}}$\\
  \midrule
    $E_\text{a}$ & 0.0237 & 0.0086 & 2.763 \\
    $E_\text{at}$ & 0.0055 & 0.0014 & 3.840 \\
    $E_\text{i}$ & 0.0144 & 0.0123 & 1.171 \\
    $E_\text{gb}$ & 0.0332 & 0.0177 & 1.874 \\
    $E_\text{gc}$ & 0.0414 & 0.0310 & 1.338 \\
    $\epsilon_\text{0}$ & 0.0016 & 0.0021 & 0.7769 \\
    $n_\text{c}$ & 0.0009 & 0.0020 & 0.4282 \\
    $c_\text{p}$ & 0.0040 & 0.0062 & 0.6490 \\
    $\sigma_\text{ts}$ & 0.0046 & 0.0011 & 4.318 \\
    $T_\text{g}$ & 0.0088 & 0.0009 & 9.969 \\
    $T_\text{m}$ & 0.0088 & 0.0018 & 4.807 \\
    $Y$ & 0.0049 & 0.0009 & 5.313 \\
    $X_\text{e}$ & 0.0214 & 0.0145 & 1.480 \\
    $X_\text{c}$ & 0.0186 & 0.0083 & 2.225 \\
    $\epsilon_\text{1.78}$ & 0.0010 & 0.0010 & 0.9407 \\
    $\epsilon_\text{15}$ & 0.0031 & 0.0018 & 1.765 \\
    $\epsilon_\text{2}$ & 0.0014 & 0.0009 & 1.631 \\
    $\epsilon_\text{3}$ & 0.0010 & 0.0005 & 2.156 \\
    $\epsilon_\text{4}$ & 0.0010 & 0.0012 & 0.8613 \\
    $\epsilon_\text{5}$ & 0.0019 & 0.0014 & 1.432 \\
    $\epsilon_\text{6}$ & 0.0010 & 0.0026 & 0.3841 \\
    $\epsilon_\text{7}$ & 0.0011 & 0.0012 & 0.8877 \\
    $\epsilon_\text{9}$ & 0.0007 & 0.0011 & 0.6105 \\
    $V_\text{ff}$ & 0.0070 & 0.0079 & 0.8951 \\
    $O_\text{i}$ & 0.0029 & 0.0027 & 1.089 \\
    $\mu_{\text{CH}_4}$ & 0.0167 & 0.0093 & 1.788 \\
    $\mu_{\text{CO}_2}$ & 0.0155 & 0.0088 & 1.764 \\
    $\mu_{\text{H}_2}$ & 0.0165 & 0.0092 & 1.795 \\
    $\mu_\text{He}$ & 0.0162 & 0.0066 & 2.454 \\
    $\mu_{\text{N}_2}$ & 0.0178 & 0.0113 & 1.581 \\
    $\mu_{\text{O}_2}$ & 0.0190 & 0.0117 & 1.630 \\
    $n_\text{e}$ & 0.0036 & 0.0021 & 1.699 \\
    $\rho$ & 0.0037 & 0.0022 & 1.690 \\
    $\delta_\text{s}$ & 0.0096 & 0.0073 & 1.330 \\
    $\lambda$ & 0.0079 & 0.0014 & 5.774 \\
    $T_\text{d}$ & 0.0123 & 0.0005 & 26.71 \\
  \bottomrule
\end{tabular}
}
\caption{The average variance of models trained with and without augmentation. The unit of each property is given in Table \ref{table:sigma}.}
\label{table:variance}
\end{table*}

Table \ref{table:variance} contains the average variance of models trained with and without augmentation, $\widehat{var}_\text{augment}$ and $\widehat{var}_\text{no augment}$, respectively, on each of the 36 properties studied in this work. We define

\begin{equation*}
    \widehat{var} = \dfrac{1}{|\mathcal{P}|} \sum_{p \in \mathcal{P}} var(
        f(x), \forall x \in \mathcal{E}_p
    )
\end{equation*}

where $\mathcal{P}$ is a set of non-equivalent repeat units, $var$ is the variance function, $f$ is a machine learning model, and $\mathcal{E}_p$ is a set of repeat units related to $p$ (a repeat unit in $\mathcal{P}$) by addition or subtraction. In this work, $\mathcal{P}$ is a set of 9 repeat units not seen by any model during training, and each $\mathcal{E}_p$ is composed of $p, 2p, 3p, 4p$ and $5p$. For example, if $p$ is \ch{(-C-)} then $\mathcal{E}_p =\{$\ch{(-C-)}, \ch{(-CC-)}, \ch{(-CCC-)}, \ch{(-CCCC-)}, \ch{(-CCCCC-)}\}.

Table \ref{table:sigma} lists standard deviations of the data in our corpus, grouped by property.

\begin{table*}[h!]
\centering
\scalebox{0.95}{
\begin{tabular}{c|c}
\toprule
Property & $\sigma$\\
\midrule
$\lambda$ & 0.0653 $\text{W/mK}$\\
$T_\text{m}$ & 109.3 $\text{K}$\\
$T_\text{d}$ & 114.7 $\text{K}$\\
$T_\text{g}$ & 109.0 $\text{K}$\\
$E_\text{at}$ & 0.470 $\text{eV}/\text{atom}$\\
$c_\text{p}$ & 0.374 $\text{J/gK}$\\
$O_\text{i}$ & 13.10 $\%$\\
$X_\text{e}$ & 18.3 $\%$\\
$V_\text{ff}$ & 0.0477 \\
$X_\text{c}$ & 23.7 $\%$\\
$\rho$ & 0.1991 $\text{g/cc}$\\
$E_\text{a}$ & 0.777 $\text{eV}$\\
$E_\text{i}$ & 1.101 $\text{eV}$\\
$E_\text{gb}$ & 1.760 $\text{eV}$\\
$E_\text{gc}$ & 1.561 $\text{eV}$\\
$\epsilon_\text{0}$ & 0.726 \\
$\epsilon_\text{1.78}$ & 1.388 \\
$\epsilon_\text{2}$ & 1.331 \\
$\epsilon_\text{3}$ & 1.276 \\
$\epsilon_\text{4}$ & 0.991 \\
$\epsilon_\text{5}$ & 1.039 \\
$\epsilon_\text{6}$ & 0.854 \\
$\epsilon_\text{15}$ & 0.359 \\
$n_\text{c}$ & 0.1713 \\
$n_\text{e}$ & 0.1142 \\
$Y$ & 1.401 $\text{MPa}$\\
$\sigma_\text{ts}$ & 40.9 $\text{MPa}$\\
$\delta_\text{s}$ & 2.64 $\sqrt \text{MPa} $\\
$\mu_\text{He}$ & 0.806 $\text{Barrer}$\\
$\mu_{\text{H}_2}$ & 0.993 $\text{Barrer}$\\
$\mu_{\text{CO}_2}$ & 1.207 $\text{Barrer}$\\
$\mu_{\text{CH}_4}$ & 1.050 $\text{Barrer}$\\
$\mu_{\text{N}_2}$ & 0.950 $\text{Barrer}$\\
$\mu_{\text{O}_2}$ & 1.018 $\text{Barrer}$\\

\bottomrule
\end{tabular}
}
\caption{The standard deviation ($\sigma$) of data in our corpus, grouped by property.}
\label{table:sigma}
\end{table*}

\end{document}